\documentclass[aps,prl,floatfix,epsfig,twocolumn,showpacs,preprintnumbers]{revtex4}

\usepackage{amssymb}
\usepackage{graphicx}
\usepackage{amsmath}

\begin{document}

\title{Calculations of Magnetic Exchange Interactions in Mott--Hubbard
Systems}
\author{Xiangang Wan$^{1,2}$, Quan Yin$^{1}$, Sergej Y. Savrasov$^{1}$}
\affiliation{$^{1}$Department of Physics, University of California, Davis, One Shields Ave, Davis, CA 95616}
\affiliation{$^{2}$National Laboratory of Solid State Microstructures and Department of
Physics, Nanjing University, Nanjing 210093, China}
\date{\today}

\begin{abstract}
An efficient method to compute magnetic exchange interactions in systems
with strong correlations is introduced. It is based on a magnetic force
theorem which evaluates linear response due to rotations of magnetic moments
and uses a novel spectral density functional framework combining our exact
diagonalization based dynamical mean field and local density functional
theories. Applications to spin waves and magnetic transition temperatures of
3\textit{d }metal mono--oxides as well as high--T$_{c}$ superconductors are
in good agreement with experiment.
\end{abstract}

\date{\today}
\maketitle

\input{epsf}

Obtaining a quantitative theory of magnetic materials spanning from
itinerant to atomic limit, above and below their temperatures of magnetic
ordering has been a theoretical challenge for many years \cite{Kubler}. By
now itinerant magnets are well described by local \textit{spin} density
approximation (L\textit{S}DA) of density functional theory (DFT) \cite{DFT},
where methodologies based on spin--spiral frozen--magnon technique \cite%
{SpinSpirals}, the use of magnetic force theorem \cite{MFT} via an
evaluation of linear response due to rotations of magnetic moments as well
as spin dynamics calculations \cite{Harmon} have allowed to access a great
variety of physical properties such as spin wave spectra , magnetic ordering
temperatures, interatomic exchange constants, dynamical susceptibilities,
etc \cite{Halilov,Tc,Bruno,Callaway,ZeitReview}.

However, there is large class of systems where calculations of exchange
interactions is still a challenging theoretical problem. These are strongly
correlated systems like high--T$_{c}$ superconductors or atomic magnets
where the on--site Coulomb interaction $U$ is comparable or larger than the
bandwidth. In cases where magnetic ordering and/or lattice distortions
result in a non--degenerate equilibrium state, techniques such as LDA+U \cite%
{LDA+Ureview} or GW\cite{GW} have been applied to describe spectroscopy,
magnetic moments, and even spin wave spectra of systems such as MnO \cite%
{LDA-MnO-dispersion}. However, in general, excitation spectra of strongly
correlated systems are not representable by single Slater determinants and
show such features as atomic multiplets \cite{Hub1}, Zhang--Rice singlets 
\cite{ZR}, Kondo resonances, etc. In Mott--Hubbard insulators the energy gap
is much larger than the magnetic ordering temperature above which a local
moment regime takes place, i.e. the system becomes paramagnetic but remains
insulating. These properties cannot be accessed either by static mean field
approaches such as L\textit{S}DA or LDA+U or by perturbation theory over the
Coulomb interaction such as GW. While versions of static \cite{SCPA} as well
as dynamic \cite{DCPA} coherent potential approximations have been
introduced in the past to access disordered local moment regime, developing
a generalized framework having a capability to compute exchange interactions
in both itinerant and atomic limits as well as in many intermediated cases
would open new opportunities in computational design of new magnetic
materials.

In the present work we develop a novel approach which is capable to deal
with this problem. Our method is based on LDA+Dynamical Mean Field Theory
(DMFT) \cite{LDA+DMFT}, a recently developed electronic structure method
describing correctly both itinerant and atomic limits and accessing ordered
and disordered moment regimes on equal footing. This is achieved by treating
correlated electrons with frequency dependent self--energies deduced from
solving corresponding Anderson impurity problem (AIM) subjected to a
self--consistency condition. Conveniently formulated using a spectral
density functional \cite{LDA+DMFT}, this LDA+DMFT method incorporates
realistic band structure effects and has already helped to solve several
long--standing problems, e.g. temperature--dependent magnetism of Fe and Ni%
\cite{DMFT-Fe-Ni}, volume collapse in Ce\cite{DMFT-Ce}, and Pu\cite{DMFT-Pu}%
, as well as electronic structure of doped Mott insulators\cite{doped-Mott}.

In order to deduce exchange constants for general wave vector $\mathbf{q}$
we utilize a linear response based magnetic force theorem \cite%
{MFT,many-body-J}and demonstrate the accuracy of our method on several
Mott--Hubbard insulators including late transition metal oxides as well as
parent materials of high--T$_{c}$ superconductors. To solve the impurity
problem which is the nucleus of the dynamical mean field algorithm we apply
a newly implemented cluster exact diagonalization method and calculate
self--consistently local Green functions, self--energies and static linear
response functions. Both the deduced spin wave spectra as well as magnetic
ordering temperatures evaluated using a Monte Carlo simulations of the
mapped Heisenberg Hamiltonians are found in good agreement with experiment.

Our implementation is based on a most recent many--body band structure
algorithm \cite{DMFT+RAT} which allows us to avoid computationally expensive
solution of the Dyson equation $[\omega -H_{0}(\mathbf{k})-\Sigma (\omega
)]G(\mathbf{k},\omega )=1$ for the electronic Green function $G(\mathbf{k}%
,\omega )$ at a large grid of frequencies $\omega .$ This is achieved by
assuming a pole expansion for the self--energy%
\begin{equation}
\Sigma (\omega )=\Sigma (\infty )+\underset{i}{\sum }V_{i}^{+}(\omega
-P_{i})^{-1}V_{i}  \label{sigma}
\end{equation}%
so that the entire problem is reduced to a matrix equation with an
energy--independent Hamiltonian

\begin{equation}
\begin{pmatrix}
\omega -H_{0}(\mathbf{k})-\Sigma (\infty ) & V^{+} \\ 
V & P%
\end{pmatrix}%
\mathcal{G}(\mathbf{k},\omega )=I  \label{gf}
\end{equation}%
where $I$ is the unit matrix and an auxiliary Green function $\mathcal{G}%
_{\alpha \beta }(\mathbf{k},\omega )$ is a matrix in the space of poles,
while the physical Green function $G(\mathbf{k},\omega )$ corresponds to the
first element of $\mathcal{G}(\mathbf{k},\omega ).$ In Eq.(\ref{sigma})
weights $V_{i}^{+},V_{i}$ and poles $P_{i}$ can be viewed as matrices which
provide a best fit to real $\Sigma (\omega ).$

It is remarkable that the present procedure allows us to use an expression
for the interatomic exchange constants similar to a linear response formula
derived within DFT \cite{MFT}. Consider second--order change in the total
energy related to the rotations of the magnetic moments appeared at sites $%
R+\tau $ and $R^{\prime }+\tau ^{\prime }$ of the lattice (here $R$ are the
lattice translations and $\tau $ are the atoms in the basis). The local
magnetic field $\mathbf{B}_{\tau }$ at every atomic site $\tau $ is
approximately described by the values of the self--energy taken at $\omega
=\infty $ [for example, $B_{\tau }^{z}=\Sigma _{\tau }^{\uparrow \uparrow
}(\infty )-\Sigma _{\tau }^{\downarrow \downarrow }(\infty )$]$.$ Thus,
according to the magnetic force theorem which assumes a rigid rotation of
atomic spin, a linear response theory expresses the interatomic exchange
constants in the form%
\begin{widetext}
\begin{equation}
J_{\tau R\tau ^{\prime }R^{\prime }}^{\alpha \beta }=\sum_{\mathbf{q}}\sum_{%
\mathbf{k}jj^{\prime }}\frac{f_{\mathbf{k}j}-f_{\mathbf{k}+\mathbf{q}%
j^{\prime }}}{\epsilon _{\mathbf{k}j}-\epsilon _{\mathbf{k}+\mathbf{q}%
j^{\prime }}}\langle \psi _{\mathbf{k}j}|[\mathbf{\sigma \times B}_{\tau
}]_{\alpha }|\psi _{\mathbf{k}+\mathbf{q}j^{\prime }}\rangle \langle \psi _{%
\mathbf{k}+\mathbf{q}j^{\prime }}|[\mathbf{\sigma \times B}_{\tau ^{\prime
}}]_{\beta }|\psi _{\mathbf{k}j}\rangle e^{i\mathbf{q}(\mathbf{R}-\mathbf{R}%
^{\prime })}  \label{jey}
\end{equation}%

\end{widetext}Here, $\sigma $ is the Pauli matrix while the one--electron
energy bands $\epsilon _{\mathbf{k}j}$ and quasiparticle wave functions $%
\psi _{\mathbf{k}j}$ appear as the solutions of the equation (\ref{gf}),
when using quasiparticle representation for the Green function $\mathcal{G}(%
\mathbf{k},\omega )$ in the form. 
\begin{equation}
\mathcal{G}(\mathbf{k},\omega )=\underset{j}{\sum }\frac{\psi _{\mathbf{k}%
j}^{+}\psi _{\mathbf{k}j}}{\omega -\epsilon _{\mathbf{k}j}}  \label{gp}
\end{equation}%
While viewed non--interacting like, this formula indeed contains major
information about many--body features in the excitation spectrum. In
particular, multiplet transitions as well as delocalized parts of the
electronic states are represented by separate "energy bands"$\epsilon _{%
\mathbf{k}j}$ including its $\mathbf{k}$ dispersion which is borrowed from
the non--interacting Hamiltonian $H_{0}(\mathbf{k})$. Thus, genuine
redistribution of spectral weight driven by the many--body interactions is
correctly captured by the present method which will give an important
feedback on the calculated exchange interactions.

There are two essential approximations which are made to make the theory
computationally tractable. As has been discussed recently \cite{many-body-J}%
, the magnetic force theorem can be introduced for a Lattinger--Ward
functional which would involve calculations of full frequency dependent
integrals between the self--energies and the Green functions. The present
method utilizes (i) the Hartree Fock values for the local magnetic fields,
and (ii)\ rational fit, i.e. Eq. (\ref{sigma}), to the self--energy, which
allows us to perform all frequency sums analytically while retaining all
major many--body multiplet features of the spectrum in the convenient linear
response expression (\ref{jey}).

To illustrate the method we consider several transition--metal oxides MnO,
FeO, CoO, NiO as well as parent high--T$_{c}$ compound CaCuO$_{2}$. All
these materials are antiferromagnetic insulators with an energy gap of a few
eV and Neel temperatures T$_{N}$ of a few hundred K. Staying below T$_{N}$
it is well known that LSDA significantly underestimates the band gap of MnO\
and NiO and fails to predict insulating character for FeO, CoO and CaCuO$%
_{2} $. The LDA+U corrects for these failures but needs to assume a symmetry
breaking for FeO and CoO. It is clear that being a Hartree Fock
approximation the LDA+U would converge to a single Slater determinant ground
state, while in many cases either degeneracy of the latter or proximity of
low--lying excited states needs to be included in statistical averagings for
the one--electron Green functions. All static mean field theories would
necessarily fail to describe paramagnetic insulating behavior. On the other
hand, the LDA+DMFT is a method valid both in ordered and local moment
regimes. Here, we consider the \textit{d} electrons of transition metal
elements as strongly correlated thus requiring dynamical treatment using
DMFT. The \textit{s} and \textit{p} electrons are assumed to be weakly
correlated and well described by the LDA Hamiltonian $H_{LDA}(\mathbf{k})$
including the full potential terms of the linear muffin-tin orbital (LMTO)
method \cite{LMTO}. To consider relativistic effects, the spin--orbit
coupling is taken into account in all cases. To obtain the one--particle
potential $H_{0}(\mathbf{k})=H_{LDA}(\mathbf{k})-V_{dc}$ entering (\ref{gf})
we subtract the double counting term $V_{dc}$ as prescribed by Ref. \cite%
{LDA+Ureview}. We use the experimental lattice structure for all materials.

In order to solve the Anderson impurity problem we implement cluster exact
diagonalization (ED) method. For transition metal mono--oxides the clusters
are chosen to include $d$ orbitals of transition metal ions hybridized with
oxygen $p$ orbitals in the octahedral environment. It has been known for
many years that such cluster exact diagonalizations provide a good
description of photoemission spectra in these materials\cite%
{Fujimori,Sawatski}. For the 2D system, such as CaCuO$_{2}$ this is reduced
to a Cu $d$ orbital surrounded by an oxygen square. This treatment allows us
to capture both an effect of atomic multiplets and of the Zhang--Rice
singlet \cite{ZR} being the lowest lying excitation of undoped high--T$_{c}$%
's. During iterations towards self--consistency, the Anderson impurity
problem is exact diagonalized each time with the positions of 3$d$ levels,
the $d$-$p$ hybridization matrix elements as well as the O $2p$ levels are
extracted from the $H_{LDA}(\mathbf{k})$. The latter being a density
functional is allowed to recompute and readjust the parameters of AIM. The
values of the Coulomb interaction $U$\ as well as the Hund's exchange $J$
were obtained earlier by the constrained LDA calculation \cite{LDA+U} and
kept fixed throughout the calculation. The AIM\ gives access to the
frequency dependent self--energies for the $d$ electrons which are then
rationally approximated assuming three poles fit in the formula (\ref{sigma}%
).

Our calculated ground state properties including magnetic moments and energy
gaps are found to be in good agreement with experiment both below and above
magnetic ordering temperatures. Below $T_{N}$ this result is in accord with
previous LDA+U\ studies \cite{LDA+U} and our numerical results indeed show
that dynamical correlations only marginally influence values of magnetic
moments in cases such as MnO, NiO and CaCuO$_{2}$. On the other hand, for
CoO and FeO, their t$_{2g}$ bands are only partially occupied and the ground
states become degenerate. Therefore small value of spin--orbital coupling
has a large effect on an appeared orbital moments which are evaluated to be
0.36 and 1.02 $\mu _{B}$ in our cluster exact diagonalized LDA+DMFT
calculations for FeO and CoO respectively. Note that the LDA+U\ would be
capable recover the insulating character only by assuming symmetry lowered
orbitally ordered solution.

We now discuss our predictions for the calculated Neel temperatures. Based
on the self--consistently obtained local Green functions and self--energies
we evaluate the interatomic exchange constants as the integral over the $%
\mathbf{q}$ space using (6,6,6) reciprocal lattice grid and utilize
Monte--Carlo simulations \cite{Tc} of the correspondingly mapped Heisenberg
Hamiltonians to find T$_{N}$. Our results are given in Table I, where
together with our most accurate exact diagonalization based simulations we
also list the predictions using LSDA\ and less accurate impurity solvers:
single atom exact diagonalization known as Hubbard I solver \cite{Hub1}, and
the Hartree--Fock approximation to atomic self--energy known as LDA+U. It is
well known that LSDA underestimates the energy gap, and as a result it
significantly overestimates the magnetic transition temperatures as seen in
Table I. (Since for FeO, CoO and CaCuO$_{2}$, LSDA converges to a completely
wrong metallic state, we omit quoting those predictions). The LDA+U method
considers the effect of $U$ in a static way, and this partially cures the
shortcomings of LSDA making $T_{N}$ smaller but still much larger than the
experimental ones. This conclusion is in agreement with the previously
calculated results for MnO \cite{LDA-MnO-dispersion}. Hubbard I\
approximation uses exact diagonalization for atomic $d$--shell and deduces
frequency dependent self--energy, which further reduces T$_{N}$ for all
transition--metal oxides. The best results are seen to be obtained by
allowing $d$--electrons to fluctuate between the bath and impurity as
prescribed by the cluster exact diagonalization calculation. This effect is
missing in the Hubbard I solver and seriously affects the electronic
structure of studied materials as, for example, for CaCuO$_{2}$ it
redistributes the spectral weight by bringing such features as the well
known Zhang--Rice singlet \cite{ZR}. In an extreme situation, where\ the
hybridization is much larger than the local Coulomb $U$, the magnetization
would eventually disappear due to strong fluctuation in the number of $d$%
--electrons at the impurity site. So it is easy to understand why the
cluster ED method gives smaller magnetic transition temperatures which are
now closer to the experiment.

It is interesting to discuss the physical reasons why T$_{N}$\ decreases
when going from NiO to MnO. These Mott--Hubbard insulators show almost
atomic values of magnetic moments $M=10-n$ corresponding to $d^{n}$
configurations, which would under assumption of the same antiferromagnetic
exchange constant $J_{AF}$ mean that the ordering temperatures should
increase with increasing the moments. However, $J_{AF}$ will decrease
significantly due to the change in the lattice parameter$.$ Also, during
evaluation of T$_{N}$ we need to account for the quantum averaging for
atomic spins directions which gives a prefactor $S(S+1)/S^{2}$ deviating
from 1 for small $S.$ To sort out these effects we first performed a sample
calculation for NiO with the expanded lattice constant of MnO$.$ The T$_{N}$
for NiO has dropped from 519K to 327K in this case. Second, since $S_{Ni}=1$
while $S_{Mn}=5/2,$ the prefactor $S(S+1)/S^{2}$ would account for a 40\%
difference so that 327$\times $1.4/2=229K is the Neel temperature that we
need to compare with our predicted T$_{N}$=172K for MnO. The residual
discrepancy can be attributed to different exchange splittings which also
affects $J_{AF}$ as pointed out earlier \cite{Terakura}.

\begin{table}[tbp]
\caption{Comparison between calculated using various approximations and
experimental magnetic transition temperatures (in K) in selected
Mott--Hubbard systems. Hubbard I\ and Cluster ED denote the results of
LDA+DMFT calculations using Hubbard I and cluster exact diagonalization
impurity solver.}%
\begin{tabular}{llllll}
\hline
& LSDA & LDA+U & Hubbard I & Cluster ED & Exp. \\ \hline
MnO & 423 & 240 & 180 & 172 & 122$^{a}$ \\ 
FeO & -- & 344 & 297 & 211 & 198$^{a}$ \\ 
CoO & -- & 407 & 356 & 300 & 291$^{a}$ \\ 
NiO & 965 & 603 & 542 & 519 & 523$^{a}$ \\ 
CaCuO$_{2}$ & -- & 765 & 698 & 602 & 537$^{b}$ \\ \hline
\end{tabular}
$^{a}$Ref. \cite{Tn-MnO-NiO}; $^{b}$Ref$.$\cite{Tn-CaCuO2}.
\end{table}

We now discuss our calculated spin--wave dispersions along major symmetry
directions in the Brillouin Zone. We illustrate this calculation on a case
of NiO for which the magnon spectra have been measured long time ago \cite%
{Spin-Wave-NiO}. Fig. 1 shows the results of our simulations using several
levels of approximations. In accord with our predictions for the Neel
temperatures, the spin waves are seriously overestimated by the LSDA theory
but get closer to the experiment once correlations are taken into account.
The best accuracy is achieved when using the cluster exact diagonalization
as the impurity solver in the LDA+DMFT\ calculation which demonstrates the
importance of many body redistribution of the spectral weight on the
calculated exchange integrals.

\begin{figure}[tbp]
\includegraphics*[height=2.5in] {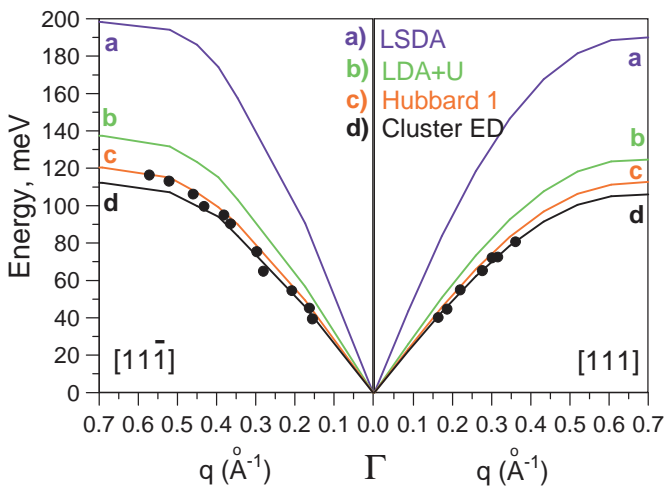}
\caption{Theoretical spin-wave dispersions for NiO calculated by LSDA,
LDA+U, Hubbard I\ and Cluster Exact Diagonalization (ED) diagonalization
impurity solvers in comparison with the experiment \protect\cite%
{Spin-Wave-NiO}.}
\label{Fig1}
\end{figure}

In summary, based on our newly implemented cluster exact diagonalization
LDA+DMFT calculation, we presented a linear response method to calculate the
exchange interaction parameters of strongly correlated systems valid as long
as mapping of total energy functional to rigid spin based Heisenberg
Hamiltonians makes sense. By using the rational interpolation for the
self--energy, our approach is very efficient, and this has allowed us to
describe quantitatively spin--wave dispersions and magnetic transition
temperatures of several realistic Mott--Hubbard insulators with many atoms
per unit cell. Applications to metallic systems are more challenging as they
may need much larger clusters to account for such subtle effects as, e.g.,
the Kondo screening, and will be carried out in the future work.

We acknowledge useful conversations with G. Kotliar and A. Lichtenstein. The
work was supported by NSF DMR grants 0608283 and 0606498.

\end{document}